\documentclass{XrU2005}
\usepackage{graphics,graphicx}

\title{New XMM-Newton Spectroscopy of the Most Luminous and Distant
  Quasars}
\author{O. Shemmer and W.~N.~Brandt}
\affil{Department of Astronomy and Astrophysics, Pennsylvania State
  University, 525 Davey Laboratory, University Park, PA 16802, USA}

\newcommand{\ltsim}{\raisebox{-.5ex}{$\;\stackrel{<}{\sim}\;$}}
\newcommand{\gtsim}{\raisebox{-.5ex}{$\;\stackrel{>}{\sim}\;$}}
\newcommand{\kms}{\ifmmode {\rm km\ s}^{-1} \else km s$^{-1}$\fi}
\newcommand{\vFWHM}{\ifmmode V_{\mbox{\tiny FWHM}} \else
            $V_{\mbox{\tiny FWHM}}$\fi}

\newcommand{\hb}{H$\beta$}
\newcommand{\mbh}{$M_{\rm BH}$}
\newcommand{\et}{et al.\ }
\newcommand{\xray}{\hbox{X-ray}}
\newcommand{\aox}{$\alpha_{\rm ox}$}
\newcommand{\nh}{$N_{\rm H}$}
\newcommand{\mb}{$M_{\rm B}$}
\newcommand{\Ka}{Fe K$\alpha$}
\newcommand{\xmm}{{\sl XMM-Newton}}
\newcommand{\chandra}{{\sl Chandra}}

\begin{document}

\maketitle

\begin{abstract}
In the two parts of this contribution we describe two related
\xmm\ programs. The first part summarizes our study of the
\xray\ spectral properties and variability of {\it z}$>$4 quasars
(Shemmer \et 2005). The second part presents preliminary results from
our ongoing \xmm\ program to investigate the \xray\ spectral
properties and variability of luminous, high accretion-rate quasars at
{\it z}$\sim$2--3. We find that the \xray\ photon index does not
depend on luminosity or redshift, and there is suggestive evidence
that it may depend on the accretion rate. None of our quasars is
significantly absorbed, and none shows signatures of reflection. By
jointly fitting high-quality spectra of eight radio-quiet {\it z}$>$4
quasars, including three from our \xmm\ observations, we place tight
constraints on the mean \xray\ spectral properties of such
sources. Most of our quasars are significantly \xray\ variable on
timescales of months--years, but none shows rapid ($\sim$1~hr
timescale) variations.
\end{abstract}

\section{\xmm\ Spectroscopy of {\it z}$>$4 Quasars}
\subsection{Introduction}
Quasars at {\it z}$>$4 are valuable cosmological probes of the
physical environment in the $\sim$1~Gyr old Universe. In particular,
the most distant quasars known, at {\it z}$\sim$6, have enabled
tracing of the physical conditions in the Universe at the end of the
re-ionization epoch with implications for large-scale structure
formation (e.g., Fan \et 2002). The study of {\it z}$>$4 quasars
therefore has become one of the main themes in astrophysics during the
past few years. One of the lines of research in this field is to
determine whether the energy production mechanism of quasars is
sensitive to the significant large-scale evolution the Universe has
experienced over cosmic time. A central question in this context is
whether black holes (BHs) in distant quasars feed and grow in the same
way as BHs in local active galactic nuclei (AGN). Recent
radio--optical observations of \hbox{{\it z}$>$4} quasars have found
that their spectral energy distributions (SEDs) are not significantly
different from those of lower redshift sources implying no SED
evolution, and hence no significant changes in the energy production
mechanism of AGN are observed (e.g., see Carilli \et 2001 and Petric
\et 2003 for radio observations; Vanden Berk \et 2001 and Pentericci
\et 2003 for \hbox{UV--optical} observations).

X-rays from distant quasars are especially valuable for studying the
energy production mechanism, since they provide information on the
innermost regions of the central engine, where most of the nuclear
energy is produced. Until fairly recently, only a handful of {\it
  z}$>$4 quasars were detected in X-rays, and the data only provided
basic \xray\ photometry. During the past five years over 100 quasars
have been detected by \chandra\ and \xmm, allowing reliable
measurements of their mean \xray\ spectral properties (e.g., Brandt
\et 2002; Bechtold \et 2003; Grupe \et 2004, 2006; Vignali \et
2003a,b, 2005). However, the different \xray\ studies of {\it z}$>$4
quasars often led to conflicting conclusions. For example, while
Bechtold \et (2003) reported that the \xray\ power-law photon indices
($\Gamma$) of {\it z}$>$4 quasars are {\em flatter} than those of
nearby AGNs, Grupe \et (2006) reported that their $\Gamma$ are rather
{\em steep}; Vignali \et (2005) found that $\Gamma$ does not undergo
significant evolution and is not luminosity dependent.

The different conclusions, frequently based upon the same \xray\ data,
were reached mainly due to the small number of photons collected in
the observations that were intended to detect {\it z}$>$4 quasars;
this led to large uncertainties in the basic \xray\ spectral
properties and hence to several possible interpretations. This
motivated us and other authors to solve the puzzle and obtain
high-quality \xray\ spectra of several \xray\ bright {\it z}$>$4
quasars. High-quality \xray\ spectra (with \gtsim500 [\gtsim100]
photons obtained by \xmm\ [\chandra]) are currently available for 10
{\it z}$>$4 radio-quiet and radio-moderate quasars (Ferrero \&
Brinkmann 2003; Farrah \et 2004; Grupe \et 2004, 2006;
Schwartz~\&~Virani~2004; Shemmer~\et~2005,~hereafter~S05). Below we
summarize the results of the recent set of five of those spectra,
which are described in detail in S05.

\subsection{High-Quality Spectra of {\it z}$>$4 Quasars}

\begin{figure}
\centering
\includegraphics[width=8.5cm]{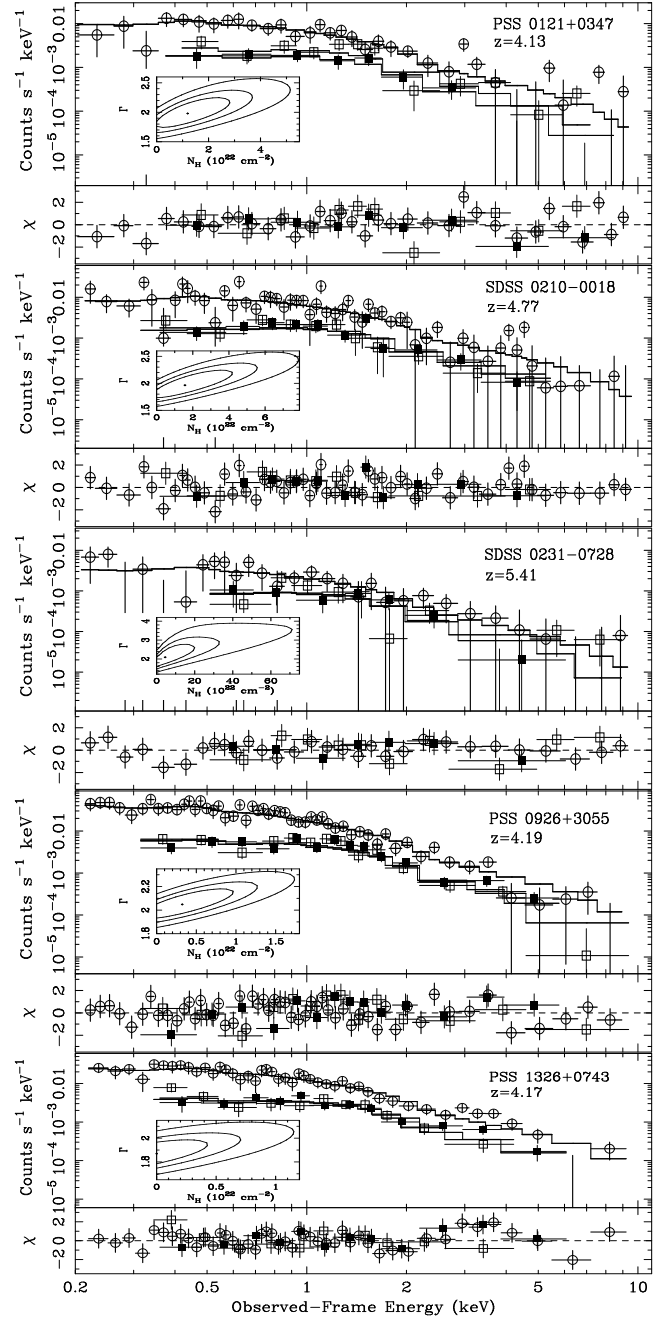}
\caption{Data, best-fit spectra, and residuals for our new
  \xmm\ observations of five \hbox{{\it z}$>$4} AGNs. Open circles,
  filled squares, and open squares represent the pn, MOS1, and MOS2
  data, respectively. Solid lines represent the best-fit model for
  each spectrum, and the thick line marks the best-fit model for the
  pn data. The $\chi$ residuals are in units of $\sigma$ with error
  bars of size 1. The inset in each panel shows 68\%, 90\%, and 99\%
  confidence contours for the intrinsic absorption and photon index.}
\label{spectra_fig}
\end{figure}

We obtained high-quality \xmm\ spectra of five {\it z}$>$4 quasars
during \xmm\ AO3; the detailed data-reduction and analysis procedures
are described in S05. Each quasar was previously detected in
\chandra\ snapshot observations (Vignali \et 2001, 2003a,b). The basic
properties of the quasars, as well as their measured
\xray\ properties, are given in Table~\ref{properties}. Three of the
quasars are radio-quiet, one quasar, PSS~0121$+$0347, is radio loud
($R$=300; Vignali \et 2003a), and another quasar, SDSS~0210$-$0018, is
radio moderate ($R$=80; Vignali \et 2001) following the radio-loudness
definition of Kellermann \et (1989). We detected $\sim$500--1500
photons from each quasar in a net exposure time of $\sim$20--30~ks per
source. These exposures enabled accurate measurements of $\Gamma$
(with $\Delta \Gamma$=0.15) and upper limits on the intrinsic neutral
column densities for each quasar. The \xmm\ data, best-fit spectra,
and residuals appear in Fig.~\ref{spectra_fig}. In
Fig.~\ref{spectra_fig} we also plot confidence contours in the
$\Gamma$--\nh\ plane for each quasar. \\

\begin{table*}
  \begin{center}
    \caption{Optical and \xray\ properties of our {\it
        z}$>$4 quasar sample.}\vspace{1em}
    \renewcommand{\arraystretch}{1.2}
    \begin{tabular}[h]{lcccccc}
      \hline
  & & & & \nh$^{\rm b}$ & $\log L_{\rm 2-10~keV}$$^{\rm a}$ & \\
Quasar & $z$ & \mb$^{\rm a}$ & $\Gamma$ & (10$^{22}$~cm$^{-2}$) &
(erg~s$^{-1}$) & \aox\ \\
      \hline
PSS 0121$+$0347 & 4.13 & $-$28.3 & $1.81^{+0.16}_{-0.16}$ & $\le2.91$
& 45.5 & $-1.65^{+0.03}_{-0.03}$ \\
SDSS 0210$-$0018 & 4.77 & $-$27.7 & $1.81^{+0.15}_{-0.14}$ & $\le4.17$
& 45.3 & $-1.54^{+0.03}_{-0.02}$ \\
SDSS 0231$-$0728 & 5.41 & $-$27.9 & $1.85^{+0.33}_{-0.31}$ & $\le19.90$
& 45.2 & $-1.62^{+0.06}_{-0.06}$ \\
PSS 0926$+$3055 & 4.19 & $-$30.1 & $1.99^{+0.08}_{-0.08}$ & $\le1.02$
& 45.9 & $-1.76^{+0.03}_{-0.01}$ \\
PSS 1326$+$0743 & 4.17 & $-$29.6 & $1.87^{+0.10}_{-0.10}$ & $\le0.47$
& 45.7 & $-1.76^{+0.03}_{-0.02}$ \\
      \hline
      \end{tabular}
\vskip 8pt
\parbox{5.4in}{
$\rm ^a${Luminosity distances were computed using the standard
    ``concordance'' cosmological parameters $\Omega_{\Lambda}$=0.7,
    $\Omega_M$=0.3, and $H_0$=70~\kms~Mpc$^{-1}$.} \\
$\rm ^b${Neutral intrinsic column density.} \\ \\ \\ \\ \\
}
    \label{properties}
  \end{center}
\end{table*}

To extend our analysis, we added to our sample high-quality
\xray\ spectra of five additional {\it z}$>$4 radio-quiet quasars
(RQQs) from the archive; these are Q~0000$-$263 (Ferrero \& Brinkmann
2003), SDSS~1030$+$0524 (Farrah \et 2004), BR~0351$-$1034 and
BR~2237$-$0607 (Grupe \et 2004, 2006), from \xmm\ observations, and
SDSS~1306$+$0356 which was observed with \chandra\ (Schwartz \& Virani
2004). The spectra of all 10 {\it z}$>$4 quasars were reduced and
analyzed uniformly to obtain the basic \xray\ spectral properties for
each source.

\subsection{\xray\ Spectral Properties of {\it z}$>$4 Radio-Quiet Quasars}

\begin{figure*}
\centering
\includegraphics[width=14cm]{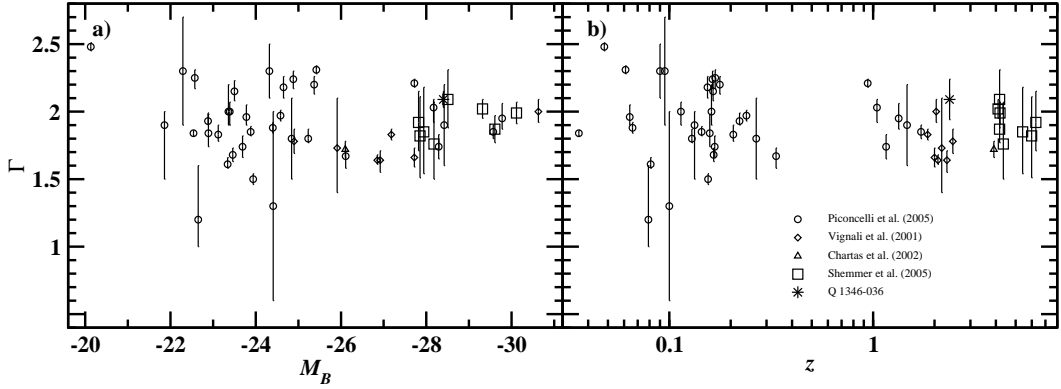}
\caption{The \xray\ photon index versus ({\it a}) absolute $B$
  magnitude and ({\it b}) redshift; adapted from S05. Note the lack of
  a clear dependence of the photon index on either luminosity or
  redshift, although considerable scatter in $\Gamma$ is observed in
  local AGN. Our new \xmm\ observation of Q~1346$-$036, a luminous,
  high accretion-rate quasar at {\it z}=2.37 is represented by a star;
  our ongoing \xmm\ observations of similar sources may determine
  whether $\Gamma$ depends on the accretion rate (see \S~2)}
\label{Gamma_MB_z}
\end{figure*}

The best-fit \xray\ spectral properties for our sources appear in
Table~\ref{properties}. The photon indices and the upper limits on the
neutral intrinsic absorption in each quasar were obtained by fitting
the spectra with intrinsically (redshifted) absorbed power-law models,
including Galactic absorption. The constraints we obtained on the
intrinsic absorption in each quasar (Table~\ref{properties}) show that
our {\it z}$>$4 RQQs are not significantly absorbed. In
Fig.~\ref{Gamma_MB_z} we plot $\Gamma$ (above 2~keV in the rest-frame)
for samples of radio-quiet AGN, including our expanded sample of {\it
  z}$>$4 quasars, against optical luminosity and redshift. We find
that $\Gamma$ takes a typical value of $\sim$1.9, and it does not
depend significantly on either optical luminosity or redshift. We also
note that there is no significant intrinsic dispersion in $\Gamma$
values within our sample of eight {\it z}$>$4 RQQs.

We have also computed optical--X-ray spectral slopes (\aox, e.g.,
Tananbaum \et 1979; see Table~\ref{properties}) for our sources and
found that our measurements are consistent with the Strateva \et
(2005) and Steffen \et (2006) conclusions that \aox\ strongly
correlates with ultraviolet luminosity and does not evolve over cosmic
time (out to {\it z}$\sim$6).

\begin{figure}
\centering
\includegraphics[width=5cm, angle=270]{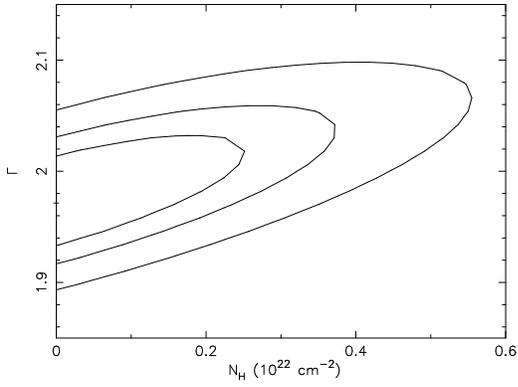}
\caption{68\%, 90\%, and 99\% confidence regions for the photon index
  vs. intrinsic column density derived from joint spectral fitting of
  our sample of eight RQQs.}
\label{joint_fit}
\end{figure}

To obtain the mean \xray\ spectral properties of the RQQ population at
{\it z}$>$4, we fitted jointly our new \xmm\ spectra of three RQQs and
the five archival high-quality spectra of {\it z}$>$4 RQQs with
several models; this is roughly equivalent to fitting a single mean
spectrum composed of $\sim$7000 photons. The number of photons in our
combined spectrum is larger by an order of magnitude than the number
of photons previously used in such analyses (e.g., Vignali \et
2005). By fitting the spectra jointly we obtained a mean photon index
$\Gamma$=1.97$^{+0.06}_{-0.04}$. We also obtained the strongest
constraint to date on the mean neutral intrinsic column density in
such sources, \nh$\ltsim$3$\times$10$^{21}$~cm$^{-2}$
(Fig.~\ref{joint_fit}), showing that optically selected RQQs at
\hbox{{\it z}$>$4} are, on average, not more absorbed than their
lower-redshift counterparts. All this suggests that the
\xray\ production mechanism and the central environment in radio-quiet
AGN have not significantly evolved over cosmic time. We also used the
combined spectrum to constrain the mean equivalent width of a putative
neutral narrow \Ka\ line to $\ltsim$190~eV, and similarly to constrain
the mean Compton-reflection component to $R\ltsim$1.2; these
constraints are consistent with the expected strength of a reflection
component given the high luminosities of our sources (e.g., Page \et
2004).

\subsection{\xray\ Variability of {\it z}$>$4 Radio-Quiet Quasars}

We applied Kolmogorov-Smirnov tests to the photon arrival times in our
new \xmm\ observations to search for rapid ($\sim$1~hr timescale in
the rest frame) variations, but none was detected.

\begin{figure}
\centering
\includegraphics[width=7cm]{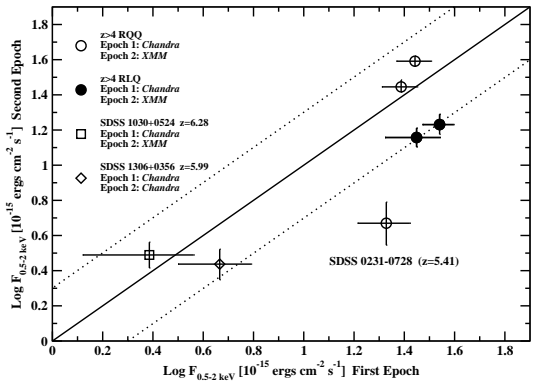}
\caption{Two-epoch Galactic-absorption corrected \hbox{0.5--2~keV}
  fluxes for seven of the \hbox{{\it z}$>$4} quasars in our
  sample. The solid line marks the 1:1 flux ratio, and the two dotted
  lines mark 1:2, and 2:1 flux ratios, to guide the
  eye. SDSS~0231$-$0728 clearly varied by more than a factor of two
  between the two epochs. The second and third most variable sources,
  PSS~0121$+$0347 and SDSS~0210$-$0018, are radio loud and radio
  moderate, respectively, and are marked with filled circles.}
\label{z4_var}
\end{figure}

To look for long-term (months--years) \xray\ variations in our sample,
we compared the fluxes of our sources in the observed-frame
\hbox{0.5--2~keV} band in the first epoch (\chandra\ snapshot
observations) with those in the second epoch (\xmm\ or
\chandra\ observations). Seven {\it z}$>$4 quasars from this study
have high-quality (i.e., \chandra\ or \xmm\ data to minimize
cross-calibration uncertainties) two-epoch \xray\ data for our
comparison. Using $\chi^2$ statistics, we found that five of the seven
quasars varied significantly between the two epochs
(Fig.~\ref{z4_var}); the two sources that did not vary significantly
between the two epochs are PSS~1326$+$0743 and SDSS~1030$+$0524.

While most quasars varied by no more than a factor of $\approx$2
between the two epochs, one source, SDSS~0231$-$0728, faded by a
factor of $\sim$4 between the first observation (\chandra) and the
second one (\xmm). This flux change occurred over a rest-frame period
of 73~d. This is the largest change in \xray\ flux observed for a
\hbox{{\it z}$>$4} RQQ. Given the UV--optical flux of the source, and
using the Strateva \et (2005) relation between UV luminosity and \aox,
it is likely that this source was caught in an \xray\ high state in
the first epoch (Vignali \et 2003b), since its \xray\ flux in the
second epoch (S05) agrees with the value predicted from its optical
flux (assuming the optical flux is nearly constant). Vignali \et
(2003b) also noted that SDSS~0231$-$0728 was \xray\ brighter than
expected (see their Fig.~5). The spectral slope of the source also
shows a possible indication of flattening from
$\Gamma$=2.8$^{+1.10}_{-0.95}$ to $\Gamma$=1.85$^{+0.33}_{-0.31}$
between the two epochs, but the significance is only $\sim$1~$\sigma$
due to the limited number of counts ($\sim$25) in the first
\chandra\ snapshot observation. This is a tentative indication for a
transition from a soft/high state to a hard/low state in this source,
as has been seen for a few local AGN (e.g., Guainazzi \et 1998;
Maccarone \et 2003).

\section{\xmm\ Spectroscopy of Luminous, High Accretion-Rate Quasars at
Redshift $\sim$2--3}
\subsection{Is $\Gamma$ an Accretion Rate Indicator?}

In \S~1 we have shown that the \xray\ photon index in RQQs appears to
be constant, with a typical value of $\sim$1.9, regardless of redshift
or luminosity. This result has also been confirmed and strengthened by
other recent studies (e.g., Mateos \et 2005; Page \et 2005; Risaliti
\& Elvis 2005). However, inspection of Fig.~\ref{Gamma_MB_z} shows
considerable scatter in $\Gamma$, in particular at low redshifts ({\it
  z}\ltsim0.5). This scatter may be attributed to a fundamental
physical parameter, which controls the \xray\ spectral shape in AGN.

Boller et al. (1996) have found that the soft ({\sl ROSAT} band)
\xray\ power-law photon index is anti-correlated with FWHM(\hb), and
hence narrow-line Seyfert 1 (NLS1s) galaxies (which meet the
FWHM[\hb]\ltsim2000~\kms\ criterion of Osterbrock \& Pogge 1985) have
significantly steeper \xray\ spectral slopes than broad-line Seyfert 1
galaxies. This trend is observed in the hard ({\sl ASCA}) \xray\ band
as well (e.g., Brandt et al. 1997; Leighly 1999). Strong correlations
between FWHM(\hb) and the \xray\ photon index in both the soft and
hard~bands are also exhibited by higher luminosity nearby ({\it
  z}\ltsim0.5) quasars (e.g., Laor \et 1997; Porquet \et 2004).

Brandt \& Boller (1998) and Laor (2000) have suggested that the strong
$\Gamma$--FWHM(\hb) correlation may be a consequence of a fundamental
correlation between $\Gamma$ and the accretion rate, since FWHM(\hb)
is considered to be an accretion-rate indicator in AGN (e.g., Boroson
\& Green 1992; Porquet \et 2004). Such a correlation may be expected
if the bulk of the emitted \hbox{optical--X-ray} energy is shifted
into higher energies for higher accretion rates.

The use of FWHM(\hb) as an accretion-rate indicator relies on
reverberation-mapping studies that found a strong correlation between
the broad-line region (BLR) size and luminosity in AGN (e.g., Kaspi
\et 2000). By assuming Keplerian motion of the BLR gas around the
central BH and using the BLR size--luminosity relation, the BH mass
becomes \mbh$=c_{1}[\lambda L_{\lambda}(5100)]^{c_{2}}$
[FWHM(\hb)]$^{2}$, and the accretion rate (in terms of the Eddington
ratio) is therefore $L/L_{\rm Edd}$$\propto$$[\lambda
  L_{\lambda}(5100)]^{1-c_{2}}$ [FWHM(\hb)]$^{-2}$, where $\lambda
L_{\lambda}(5100)$ is the monochromatic luminosity at 5100\AA, $L_{\rm
  Edd}$ is the Eddington luminosity, and $c_{1}$ and $c_{2}$ are
constants determined by reverberation mapping (e.g., Kaspi et
al. 2000, 2005; see the specific equations in Shemmer \et 2004,
hereafter S04).  FWHM(\hb) is perhaps the best accretion-rate
indicator, and the use of other emission lines as proxies to \hb, such
as C~{\sc iv}, can lead to spurious estimates of $L/L_{\rm Edd}$
(e.g., Baskin \& Laor 2005). NLS1s are the highest accretion-rate
sources among low-luminosity AGN, with $L/L_{\rm Edd}$ approaching,
and in extreme cases even exceeding, unity.

\begin{figure}
\centering
\includegraphics[width=8cm]{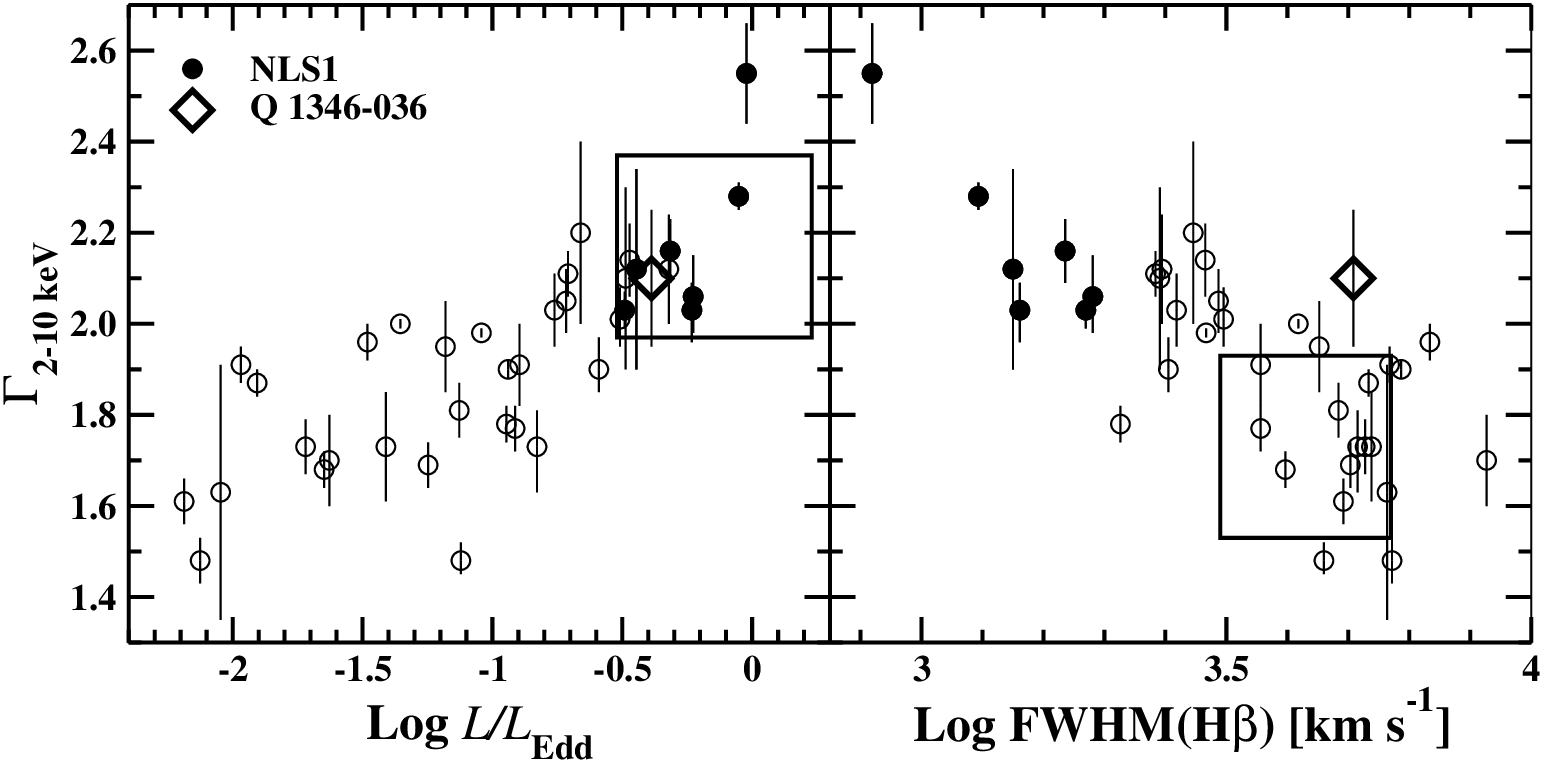}
\caption{\xray\ photon index in the rest-frame \hbox{2--10~keV} band
  versus $L/L_{\rm Edd}$ ({\it left}) and FWHM(\hb) ({\it
    right}). Circles mark AGN at {\it z}$<$0.5. NLS1s are marked with
  filled symbols, and Q~1346$-$036, a luminous, {\it z}=2.37 quasar
  from the S04 sample and recently observed by \xmm, is marked with a
  diamond; it is the only high-$z$ source on this diagram. Both
  $L/L_{\rm Edd}$ and FWHM(\hb) are significantly correlated with
  $\Gamma$ for {\it z}$<$0.5 AGN. Boxes mark the expected positions of
  the S04 quasars on each correlation.}
\label{GE}
\end{figure}

\subsection{\xray\ Properties of Luminous, High Accretion Rate Quasars at
High Redshift}

The recent study of S04 has found that in at least two respects,
accretion rate (determined from \hb) and metallicity, extremely
luminous ({\it L}\gtsim10$^{47}$~erg~s$^{-1}$, where $L$ is the
bolometric luminosity) quasars at 2$<${\it z}$<$3.5 resemble NLS1s
with {\it L}\ltsim$10^{45}$~erg~s$^{-1}$.  Motivated by this study, we
have initiated an \xmm\ program to look for unusual \xray\ properties
in the S04 quasars and to determine whether they are the luminous,
high-$z$ analogs of local NLS1s (frequently termed narrow-line type~1
quasars). Specifically, we intend to measure $\Gamma$ accurately (to
within $\pm$0.15) for these sources and to look for rapid (on a
$\sim$1~hr rest-frame timescale) \xray\ variations, since steep photon
indices and rapid \xray\ variations are two well-known NLS1
characteristics. The \xray\ photon indices obtained for the S04
sources will allow us to test whether $\Gamma$ can be considered a
reliable accretion-rate indicator for all AGN, including luminous,
high-$z$ quasars, with important implications for accretion disk and
corona models in AGN (e.g., Haardt \& Maraschi 1993).

\begin{figure}
\centering
\includegraphics[width=8cm]{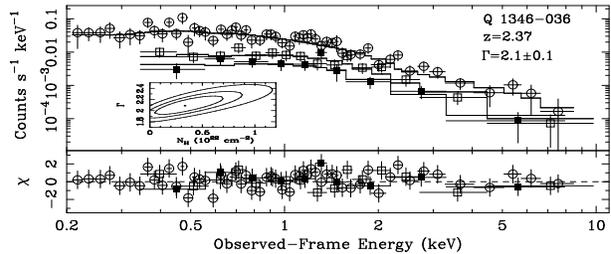}
\caption{Data, best-fit spectrum, and residuals for our new
  \xmm\ observation of Q~1346$-$036, a luminous {\it z}=2.37 quasar
  from S04. Symbols are similar to those in Fig.~\ref{spectra_fig}. We
  find a photon index $\Gamma$=2.1$\pm$0.1, which seems to support the
  hypothesis that the accretion rate is the underlying physical driver
  for steep \xray\ spectra in AGN (see Fig.~\ref{GE}).}
\label{q1346}
\end{figure}

This test is portrayed in Fig.~\ref{GE}, where we have plotted
$\Gamma$ versus FWHM(\hb) and $L/L_{\rm Edd}$ (which is a combination
of FWHM[\hb] and $L$). In this plot we consider archival data for AGN
with high-quality \xray\ spectra (obtained from Reeves \et 1997;
Reynolds 1997; George \et 1998, 2000; Piconcelli \et 2005) and with
reliable FWHM(\hb) measurements. All but one of the sources in
Fig.~\ref{GE}, Q~1346$-$036, are AGN at {\it z}\ltsim0.5 (and
therefore have low--moderate luminosities), since at higher redshift
\hb\ is not present in the optical band and near-IR measurements of
FWHM(\hb) are difficult to obtain. Although there are significant
correlations between $\Gamma$ and both FWHM(\hb) and $L/L_{\rm Edd}$
for the nearby sources, the S04 quasars are predicted to have
significantly different values of $\Gamma$ in each case. Based on
their FWHM(\hb), these quasars are expected to have a mean $\Gamma$ of
$\sim$1.7, but when their high accretion rates are considered, the
mean expected $\Gamma$ is $\sim$2.2, which is only observed in extreme
NLS1s. The first S04 quasar observed in our ongoing \xmm\ program,
Q~1346$-$036, shows a moderately steep \xray\ spectrum (see
Fig.~\ref{q1346}) and suggests that the $\Gamma$--$L/L_{\rm Edd}$
correlation may still hold when luminous, high-$z$ quasars are
included (Fig.~\ref{GE}).

\section*{Acknowledgments}

We wish to thank all of our collaborators who contributed to these
projects: X.~Fan, S.~Kaspi, R.~Maiolino, H.~Netzer, G.~T.~Richards,
D.~P.~Schneider, M.~A.~Strauss, and C.~Vignali.

\end{document}